\documentclass[sigconf]{acmart}
\acmConference[EASE 2025]{The 29th International Conference on Evaluation and Assessment in Software Engineering}{17–20 June, 2025}{Istanbul, Türkiye}

\usepackage{enumitem}
\usepackage{listings}
\usepackage{multirow}
\usepackage{makecell}
\usepackage{subcaption}
\usepackage{tcolorbox}
\usepackage{xspace}

\newcommand{\etal}{\emph{et al.}\xspace}

\begin{document}

\title{The Impact of Environment Configurations on the Stability of AI-Enabled Systems}

\author{Musfiqur Rahman}
\affiliation{%
    \institution{Concordia University}
    \city{Montr\'eal}
    \country{Canada}
}
\email{musfiqur.rahman@mail.concordia.ca}

\author{SayedHassan Khatoonabadi}
\affiliation{%
    \institution{Concordia University}
    \city{Montr\'eal}
    \country{Canada}
}
\email{sayedhassan.khatoonabadi@concordia.ca}

\author{Ahmad Abdellatif}
\affiliation{%
    \institution{University of Calgary}
    \city{Calgary}
    \country{Canada}
}
\email{ahmad.abdellatif@ucalgary.ca}

\author{Haya Samaana}
\affiliation{%
    \institution{An-Najah National University}
    \city{Nablus}
    \country{Palestine}
}
\email{hayasam@najah.edu}

\author{Emad Shihab}
\affiliation{%
    \institution{Concordia University}
    \city{Montr\'eal}
    \country{Canada}
}
\email{emad.shihab@concordia.ca}

\renewcommand{\shortauthors}{Rahman et al.}

\begin{abstract}
    Nowadays, software systems tend to include Artificial Intelligence (AI) components. Changes in the operational environment have been known to negatively impact the stability of AI-enabled software systems by causing unintended changes in behavior. However, how an environment configuration impacts the behavior of such systems has yet to be explored. Understanding and quantifying the degree of instability caused by different environment settings can help practitioners decide the best environment configuration for the most stable AI systems. To achieve this goal, we performed experiments with eight different combinations of three key environment variables (operating system, Python version, and CPU architecture) on $30$ open-source AI-enabled systems using the Travis CI platform. We determine the existence and the degree of instability introduced by each configuration using three metrics: the output of an AI component of the system (model performance), the time required to build and run the system (processing time), and the cost associated with building and running the system (expense). Our results indicate that changes in environment configurations lead to instability across all three metrics; however, it is observed more frequently with respect to processing time and expense rather than model performance. For example, between Linux and MacOS, instability is observed in 23\%, 96.67\%, and 100\% of the studied projects in model performance, processing time, and expense, respectively. Our findings underscore the importance of identifying the optimal combination of configuration settings to mitigate drops in model performance and reduce the processing time and expense before deploying an AI-enabled system.
\end{abstract}




\maketitle

\section{Introduction}
With the recent advances and popularity in the field of Artificial Intelligence (AI)---more specifically Machine Learning (ML) models---in solving numerous real-life problems, more and more software systems are integrating such models as part of their pipeline~\cite{martinez2022software}. Software systems are inherently complex and the issue of stability in software systems has been previously investigated~\cite{bevan2003identification,maisikeli2018measuring}. In addition, any ML model is, at its core, probabilistic and, as a result, suffers from uncertainty.~\cite{klas2018uncertainty,kompa2021second,hammer2007process}. The complexity of a software system, combined with the nondeterministic nature of an ML model, can make AI-enabled systems behave inconsistently across different operational environments. In this study, the term `operational environment' refers to any environment where an AI-enabled system is built and/or served, such as development and deployment environments. This inconsistent behavior introduces instability---the phenomenon where a piece of software behaves differently when the operational environment changes, even though the internal software artifacts, such as code and input data, remain the same. Such instability indicates low adaptability~\cite{iso} of a system, which is undesirable system behavior.

In practice, development and deployment environments may differ. The potential for a substantial difference between development and deployment behavior, termed \textit{training/serving skew}, has been reported before~\cite{bogner2021characterizing,sculley2015hidden}. For example, an arbitrary face recognition system that achieves an F1 score of, say $0.9$, in the development environment may not achieve a similar F1 score once deployed in a different environment configuration. Therefore, understanding how an ML model may behave differently after deployment compared to its behavior in the development environment is a crucial aspect of AI-enabled software development. Although the literature has previously investigated the instability in the behavior of ML models from the ML algorithm perspective~\cite{summers2021nondeterminism,kedziora2024prediction}, the developers must also determine the degree of instability that can be introduced due to environment configurations as well. Therefore, running the system under different configuration settings should be an additional step before deployment of the system to determine whether model performance varies significantly across configurations. As demonstrated by the previous example, the probabilistic and uncertain nature of ML models can introduce novel challenges affecting different stages of the software development life cycle. The software engineering research community has recently begun investigating the challenges associated with the uncertain nature of AI-enabled software systems. These challenges affect various aspects of the development life cycle, including requirement elicitation~\cite{belani2019requirements}, software testing and quality assurance~\cite{felderer2021quality}, and deployment~\cite{john2021architecting}.

As discussed above, the environment settings can vary from one stage to another in the development life cycle. The choices made by the developers regarding development environment settings, such as operating systems, versions of a programming language, and hardware configurations, can depend on many factors, including developers' experience, business needs, and existing environment configurations of legacy systems. However, these choices may potentially introduce instability in the prediction quality of AI/ML models as ``practitioners' degrees of freedom''~\cite{wicherts2016degrees,simmons2011false}, which is a known issue in the field of applied statistics. However, in the domain of software engineering, there is no existing work studying the potential sources of instability in AI-enabled software from an environment configuration perspective. We use the term `instability' as a quantitative measure throughout the study to assess the extent to which an AI-enabled system behaves differently when environment configurations change. We aim to achieve this goal by experimenting with eight combinations of three environment variables, namely operating system, Python version, and CPU architecture. We conduct experiments on $30$ open-source AI-enabled projects using the Travis CI platform, measuring instability in model performance, processing time, and expense. Specifically, we aim to answer the following three research questions:

%\begin{itemize}
    \textbf{RQ1: (Operating System) How much instability is introduced by changing the operating system in AI-enabled systems?} We analyze whether variations in operating systems make AI-enabled systems behave differently. We observed instability in model performance across 23\% of the projects between Linux and MacOS whereas 20\% of the projects show such instability between Linux and Windows. Almost all projects show significant instability in processing time and expense between different operating systems.
    
    \textbf{RQ2: (Python Version) How much does changing the Python version introduce instability in AI-enabled systems?} Python is the most frequently used programming language for AI-enabled systems. Therefore, it is critical to investigate the effect of Python versions on the behavior of AI-enabled systems. We found that Python 3.6 and Python 3.7 consistently produce identical results in all three metrics. However, between Python 3.7 and Python 3.8, instability can be observed in about 17\% of the projects in model performance and 80\% of the projects in both processing time and expense.
    
    \textbf{RQ3: (CPU Architecture) How much does changing the CPU architecture introduce instability in AI-enabled systems?} We turn our focus from software-level configuration to hardware-level configuration. We compare two CPU architectures and find that over 93\% of the projects show instability in processing time and expense while only 20\% of them vary in model performance between AMD64 and ARM64 architectures.
%\end{itemize}

Our findings imply that changes in configuration settings are very likely to introduce significant instability in AI-enabled systems although the degree varies from project to project. Significant instability in AI-enabled systems in processing time and expense is more frequently observed than model performance. Determining the best configuration settings for a project is an iterative process, and developers should build and run their systems on different settings to find the most optimized environment configuration for the system.

In summary, this paper makes the following contributions:

\begin{itemize}[leftmargin=*]
    \item To the best of our knowledge, this is the first empirical study on the instability of AI-enabled systems from the environment configuration point of view.
    \item We provide empirical evidence behind the necessity of \textit{dev/prod parity} principle where development and production environments are kept similar as much as possible~\cite{hoffman2016beyond}.
    \item We make our data and scripts available for reproducibility and future research~\cite{replication}.
\end{itemize}

The rest of the paper is organized as follows. Section~\ref{methodology} covers background and methodology. Sections~\ref{rq1}–\ref{rq3} present findings for each research question. Sections~\ref{discussion} and~\ref{threats} discuss our results and threats to validity. Section~\ref{related-work} reviews related work, and Section~\ref{conclusion} concludes with a summary and future directions.

\section{Methodology and Background} \label{methodology}
We use Travis CI—a widely used Continuous Integration (CI) platform~\cite{hilton2016usage}—to run experiments across different operational configurations. We chose Travis CI because it’s the most popular CI tool among OSS developers for AI-enabled systems~\cite{rzig2022characterizing}.

\subsection{Environment Configurations in Travis CI} \label{env_conf}
In this study, the three configuration variables we experiment with are \emph{Operating System}, \emph{CPU Architecture}, and \emph{Python Version}. We choose to experiment with these three variables because the operating system is the core of any development environment where a system is built and run, the CPU is the core of the hardware on which a system is run, and the programming language is at the core of development tech stack used for building a system. In our experiment, we use the following list of options for each configuration variable:

%\begin{itemize}[leftmargin=*]
    \emph{\textbf{Operating System:}} Linux (version Ubuntu-Xenial 16.04), MacOS (version 10.14.4), and Windows (10 version 1803). We chose these three operating systems because they are the most common operating systems used in development stacks across the globe~\cite{DesktopO81:online}. Within Linux, we experiment with three different distributions, which are Ubuntu-Xenial 16.04, Ubuntu-Bionic 18.04, and Ubuntu-Focal 20.04. We chose these three distributions because, during the time we were running our experiments, these three distributions were the latest Long Term Support (LTS) versions of the top three most recent Ubuntu distributions. For brevity, we will only use the distribution name throughout the rest of the paper.

    \emph{\textbf{Python Version:}} 3.6, 3.7, and 3.8. We chose these three versions because, during the time of our experiments, Python 3.7 was the oldest version of Python that was being maintained~\cite{Statusof80:online}. Furthermore, the majority of the projects in our dataset were developed using Python 3.7 or older versions. We compare Python 3.7 against one earlier (Python 3.6) and one later version (Python 3.8) so that features of different versions are still comparable and not significantly different from one another.
    
    \emph{\textbf{CPU Architecture:}} AMD64 and ARM64. We chose these two architectures because they are commonly compared against each other from a variety of points of view~\cite{blem2013power,bhandarkar1997risc,sankaralingam2013detailed}. Furthermore, a recent study shows that ARM64 architecture is considered an alternative to traditional AMD64 architectures, which is gaining interest among software developers~\cite{chen2023x86}.
%\end{itemize}

We compare all configuration settings against a baseline configuration to quantify the instability. The baseline configuration is defined as Linux with Xenial distribution for the operating system, AMD64 for the CPU architecture, and Python 3.7 for the programming language. The reason behind this choice is that these were the default values set by Travis CI at the time of conducting the experiments. It is important to note that there is always one and only one environment variable that is different from the baseline configuration. We apply this condition to make sure that if there is instability, it is due to the variable that is different from the baseline and nothing else. A total of seven environment configurations are selected to be compared with the baseline configuration as shown in Table~\ref{env_conf_tbl}.

\begin{table*}
\centering
\caption{\label{env_conf_tbl} Environment configurations compared against the baseline configuration: \texttt{os:linux}, \texttt{dist:xenial}, \texttt{arch:amd64}, \texttt{python:3.7}. In each row, the variable that is different from the baseline is \underline{underlined}.}
\begin{tabular}{|ccc|c|}
\hline
\multicolumn{1}{|c|}{\textbf{Purpose}} &
  \multicolumn{1}{c|}{\textbf{Comparison}} &
  \textbf{Configuration} &
  \textbf{\makecell{Total \\ Configurations}} \\ \hline
\multicolumn{1}{|c|}{\multirow{2}{*}{\textbf{Effect of operating system}}} &
  \multicolumn{1}{c|}{\textit{Linux vs MacOS}} &
  \underline{\texttt{os:osx}}, \texttt{arch:amd64}, \texttt{python:3.7} &
  \multirow{2}{*}{2} \\ \cline{2-3}
\multicolumn{1}{|c|}{} &
  \multicolumn{1}{c|}{\textit{Linux vs Windows}} &
  \underline{\texttt{os:windows}}, \texttt{arch:amd64}, \texttt{python:3.7} &
   \\ \hline
\multicolumn{1}{|c|}{\multirow{2}{*}{\textbf{Effect of distribution}}} &
  \multicolumn{1}{c|}{\textit{Linux-Xenial vs Linux-Bionic}} &
  \texttt{os:linux}, \underline{\tt{dist:bionic}}, \texttt{arch:amd64}, \texttt{python:3.7} &
  \multirow{2}{*}{2} \\ \cline{2-3}
\multicolumn{1}{|c|}{} &
  \multicolumn{1}{c|}{\textit{Linux-Xenial vs Linux-Focal}} &
  \texttt{os:linux}, \underline{\texttt{dist:focal}}, \texttt{arch:amd64}, \texttt{python:3.7} &
   \\ \hline
\multicolumn{1}{|c|}{\multirow{2}{*}{\textbf{Effect of Python version}}} &
  \multicolumn{1}{c|}{\textit{Python 3.6 vs Python 3.7}} &
  \texttt{os:linux}, \texttt{arch:amd64}, \underline{\texttt{python:3.6}} &
  \multirow{2}{*}{2} \\ \cline{2-3}
\multicolumn{1}{|c|}{} &
  \multicolumn{1}{c|}{\textit{Python 3.7 vs Python 3.8}} &
  \texttt{os:linux}, \texttt{arch:amd64}, \underline{\texttt{python:3.8}} &
   \\ \hline
\multicolumn{1}{|c|}{\textbf{Effect of CPU architecture}} &
  \multicolumn{1}{c|}{\textit{AMD64 vs ARM64}} &
  \texttt{os:linux}, \underline{\texttt{arch:arm64}}, \texttt{python:3.7} &
  1 \\ \hline
\end{tabular}
\end{table*}

The configurations are defined in a \texttt{.travis.yml} file which is written in YAML-based Domain Specific Language~\cite{domain_specific_language_2024}. An example of a typical \texttt{.travis.yml} file is shown in Listing~\ref{example_travis_yml} which defines a \textit{build} with two \textit{jobs} each of which has three \textit{phases}.

\begin{lstlisting}[language = XML, numbers=left, caption=An example of a typical \texttt{.travis.yml} file, captionpos=b, basicstyle=\ttfamily\small, breaklines, keywordstyle=\color{red}, tabsize=1, label=example_travis_yml, showspaces=false]
language: python
jobs:
  include:
    - name: Python 3.6 on Linux-Xenial
      python: 3.6
      os: linux
      dist: xenial
      arch: arm64
    - name: Python 3.7 on Linux-Bionic
      python: 3.7
      os: linux
      dist: bionic
      arch: amd64
install:
  - pip3 install --upgrade pip
  - pip3 install -r requirements.txt
script:
  - python3 src/train.py
  - python3 src/test.py
after_success:
  - echo "Successful."
\end{lstlisting}

We define the three key Travis CI terminologies below:

%\begin{itemize}[leftmargin=*]
    \emph{\textbf{Job:}} A \textit{job} is defined as an automated process that clones a repository into a virtual environment (VM). A \textit{job} carries out a series of \textit{phases}.
    
    \emph{\textbf{Phase:}} A \textit{phase} is one sequential step of a \textit{job}. There are two main Travis CI \textit{phases}, namely, \texttt{install} and \texttt{script}. Installation of any dependencies required to build a software project is performed in the \texttt{install} \textit{phase} whereas the \texttt{script} \textit{phase} runs the build scripts. Travis CI also supports three optional deployment \textit{phases}: \texttt{before\_deploy}, \texttt{deploy}, and \texttt{after\_deploy}. Custom commands like \texttt{after\_success} and \texttt{after\_failure} can also be run as part of a \textit{phase}.
    
    \emph{\textbf{Build:}} A \textit{build} is a group of \textit{jobs}. By default, \textit{jobs} in a build run in sequence, although depending on one's subscription plan, \textit{jobs} can be run concurrently.
%\end{itemize}

The configuration settings of a VM are described using a set of keywords. For instance, \texttt{OS} and \texttt{language} are two configuration-related keywords. \texttt{OS} sets the Operating System of a VM for a particular \textit{job} whereas the \texttt{language} keyword is used to prepare the VM by setting up tools of a specific programming language. In Listing~\ref{example_travis_yml}, Python is set as the programming language for the project in line 1. Line 2 marks the beginning of the \texttt{jobs} block. In this example, two independent \textit{jobs} are defined. The first \textit{job} (line 4--8) will run on a VM with an ARM64 CPU and Linux-Xenial distribution as the operating system. Python version 3.6 is installed to run the Python scripts. Similarly, the second \textit{job} (line 9--13) will run on a VM where the operating system is Linux-Bionic and the CPU architecture is AMD64. Python version 3.7 is used to run the Python scripts. In both \textit{jobs}, after the VMs are spun up, \texttt{pip} is upgraded (line 15) and required libraries are installed (line 16). Once all the dependencies are installed, two Python scripts from the \texttt{src} folder are run sequentially: \texttt{train.py} (line 18) and \texttt{test.py} (line 19). After the successful execution of the \texttt{script} \textit{phase}, a message ``Successful.'' is displayed on the screen (line 21).

\subsection{Dataset} \label{data}
In this study, we use a dataset of open-source AI-enabled projects from GitHub curated by Rzig~\etal~\cite{rzig2022characterizing}. We chose this dataset because all these projects use Travis CI and are primarily written in Python. We focus on Python projects only because it has been reported that Python is the most popular programming language for the development of AI and ML-based solutions~\cite{sultonov2023importance,raschka2020machine,gonzalez2020state}. We clone all projects and build them in the Travis CI platform using the existing \texttt{.travis.yml} files. Once built, a project has one of the following statuses:

%\begin{itemize}[leftmargin=*]
    \textit{\textbf{Errored:}} An \textit{errored} \textit{build} has one or more \textit{errored} \textit{job(s)}. A \textit{job} that encounters an issue during the \texttt{install} \textit{phase} receives the \textit{errored} status.
    
    \textit{\textbf{Failed:}} A \textit{failed} \textit{build} has one or more \textit{failed} \textit{job(s)}. A \textit{job} that encounters an issue during the \texttt{script} \textit{phase} receives the \textit{failed} status.
    
    \textit{\textbf{Passed:}} A \textit{build} receives the \textit{passed} status when all \textit{jobs} receive the \textit{passed} status.
%\end{itemize}

Since the goal of our work is to study instability in these projects, it is required that all projects are successfully built under all configuration settings described in Section~\ref{env_conf}. For example, if a project only runs on Linux, but not on MacOS and/or Windows, then we cannot quantify instability due to the change in operating system in this project. However, the majority of the projects are not developed with the aim of running them on all major operating systems, CPU architectures, or multiple versions of Python. For example, fer~\cite{fer2018} is one of the projects in the dataset. The \texttt{.travis.yml} file in this project shows that it was developed for and tested on Linux-Xenial, AMD64 CPU architecture, and Python 3.6. While we tried to edit the \texttt{.travis.yml} files of all the projects in the dataset to incorporate all the configuration settings from Section~\ref{env_conf}, for the majority of the projects we were unsuccessful in building them in all those settings because open-source projects usually are developed and tested on a small subset of many possible configuration settings. For example, several projects were built using Python 3.6 and when we tried to build them with Python 3.8 they failed due to dependency issues and version mismatch between Python libraries. $30$ projects returned a \textit{build} status of \texttt{passed} under all configuration settings under investigation. We move forward with these $30$ projects for further analysis. Table~\ref{tab:project-overview} provides an overview of the $30$ projects used in this study. The full list of studied projects can be found in the replication package~\cite{replication}. Building and running $30$ projects on Travis CI took a total of $1185.87$ build hours and cost us $1,566,775$ build credits which is worth $\$940$ excluding the monthly subscription fee of $\$260$.

\begin{table}
\centering
\caption{\label{tab:project-overview} Overview of the $30$ projects used in this study.}
\begin{tabular}{c|c|c|c|c|c|}
\cline{2-6}
\textbf{}                                   & \textbf{Avg.} & \textbf{Std.} & \textbf{Min.} & \textbf{Med.} & \textbf{Max.} \\ \hline
\multicolumn{1}{|c|}{\textbf{Commits}}      & 437.33        & 393.24        & 13.00         & 331.50        & 1570.00       \\ \hline
\multicolumn{1}{|c|}{\textbf{Forks}}        & 84.17         & 122.10        & 6.00          & 43.00         & 548.00        \\ \hline
\multicolumn{1}{|c|}{\textbf{Stars}}        & 402.90        & 690.30        & 9.00          & 179.50        & 2949.00       \\ \hline
\multicolumn{1}{|c|}{\textbf{Contributors}} & 8.70          & 12.22         & 1.00          & 5.00          & 67.00         \\ \hline
\end{tabular}
\end{table}
    
\subsection{Analysis of Instability} \label{var_analysis}
\subsubsection{Evaluation Metrics:}
One of the reasons why the popularity of AI-enabled systems has shown consistent growth over the last few years is that these systems are becoming more and more accurate in solving real-life problems. With the increasing amount of high-quality data, these systems are expected to perform better over time~\cite{domingos2012few}. Therefore, the primary factor that determines if an AI-enabled system is practically useful or not is how well it performs in accomplishing a given task. The secondary factor that influences a system's practical usefulness is whether it can accomplish a task within a reasonable amount of time. This implies that like any other traditional software system, both model performance and time are critical aspects of an AI-enabled system as well. However, that is not all. Because AI-enabled systems are trained on existing data, any changes in the data cause model performance degradation over time~\cite{nelson2015evaluating}. This necessitates frequent retraining of a system within a reasonable amount of time. Research in less time-consuming training of AI-enabled systems has been an interesting topic for a while~\cite{kavikondala2019automated,wu2020deltagrad,mahadevan2024cost,kim2022efficient}. To further facilitate this process of frequent improvement of a system by retraining it, many online cloud platforms offer paid services that can be utilized. These services provide users with different computation resources such as high volumes of RAM, GPUs, and TPUs. Of course, these services are not free and usually, a user needs to pay at an hourly rate~\cite{EC2OnDem9:online,VMinstan1:online}. This brings in the third most important factor which is the expense associated with building and running an AI-enabled system. In our investigation of instability, we therefore pay attention to these three factors as discussed below:

%\begin{itemize}[leftmargin=*]
 \emph{\textbf{Model performance:}} This metric is determined from the model performance of the AI component of the system. For each project, we create a Python script named \texttt{example.py}. In this script, we implement an example use case of respective projects. Some projects, such as StarBoost~\cite{halford2018starboost}, already have example scripts and/or notebooks that demo one or more key use cases of those projects. In other projects where no example scripts/notebooks are available (such as PyALCS~\cite{kozlowski2018pyalcs}), we go through the tutorial sections of their documentation and find example use cases. This is a crucial step in our experimental setup because the \texttt{example.py} scripts define and run ML tasks like regression and classification. The outputs of these scripts are some numeric measures like \emph{F1-score} (for classification) and \emph{R\textsuperscript{2}} (for regression). This numeric measure is the model performance-related metric.
    
 \emph{\textbf{Processing time:}} This metric is obtained from the total processing time (in minutes) taken to run a project in a given environment configuration. In other words, it is the time taken to complete a \emph{job} in Travis CI. This includes spinning up the VM, installing required libraries and modules, building the project, and running the \texttt{example.py} script.
    
 \emph{\textbf{Expense:}} This metric is obtained from the amount of Travis CI credits spent on building and running the example script for each project. The number of credits associated with processing a project in Travis CI is calculated based on the amount of time it takes from spinning up the VM to executing the last \emph{phase} in the \texttt{.travis.yml} file. In other words, the longer it takes to complete processing a project, the more credits are spent. The number of credits required to run a project on a VM in the Travis CI environment is determined only by the operating system of the VM and nothing else. This means that credits are deducted at different rates only when operating systems are different. The billing documentation from the official Travis CI website~\cite{travisci_documentation_2024} states that the number of credits spent per minute on running a VM with Linux, Windows, and MacOS are respectively $10$, $20$, and $50$. We realize that processing time and expense are correlated and it may seem redundant to study expense as a separate metric. However, the scale of processing time and expense can be considerably different. Let us take an arbitrary example. If a project takes $120$ minutes to complete on Linux and $121$ minutes to complete on MacOS, the processing time differs only by one unit, and a one-unit difference may not be significant. However, when we consider the number of credits spent, these values are $120\times10=1200$ and $121\times50=6050$ for Linux and MacOS, respectively. When we convert the number of credits to the equivalent dollar amounts at a rate of $0.0006$ dollars per credit as calculated from~\cite{travisci_documentation_2024}, they are $1200\times0.0006=0.72$ and $6050\times0.0006=3.63$ for Linux and MacOS, respectively. As this example demonstrates, there can be scenarios where the difference in processing time between different settings is small and insignificant, however, the difference in cost can still be big and significant. This is why we study processing time and expense as two separate metrics in this study.
%\end{itemize}

\subsubsection{Result Analysis:}
We run each project $50$ times under each configuration shown in Table~\ref{env_conf_tbl}. The purpose behind choosing to generate a distribution of $50$ runs per configuration per project is to mitigate random and unaccounted-for fluctuations in the metrics. For example, let us assume that we aim to determine how the \emph{model performance} of a project varies due to CPU architecture. In this case, we generate a distribution of \emph{model performance} for the project by running it $50$ times under the configuration of \texttt{os:linux}, \texttt{dist:xenial}, \underline{\texttt{arch:arm64}} and \texttt{python:3.7}. This distribution is then compared against the distribution of \emph{model performance} generated from $50$ runs of the same project under the baseline configuration which is \texttt{os:linux}, \texttt{dist:xenial}, \underline{\texttt{arch:amd64}} and \texttt{python:3.7}. As previously mentioned, there is always one and only one environment variable that is different from the baseline configuration. In this example, the only variable that is different from the baseline configuration is \texttt{arch}, as underlined above. Once these two distributions are generated, we then perform the following steps:

\emph{Step 1}: For each project, we calculate the percentage change as follows:

\begin{equation} \label{eq:1}
    \begin{split}
        P = \frac{\overline{m_o}-\overline{m_b}}{\mid\overline{m_b}\mid}\times100
    \end{split}
\end{equation}

The variables in the Equation~\ref{eq:1} are defined as follows:

\begin{itemize}[leftmargin=*]
    \item $\overline{m_b}$ is the arithmetic mean of a metric (\textit{model performance}, \textit{processing time}, or \textit{expense}) obtained from $50$ runs of a project under the baseline configuration.
    \item $\overline{m_o}$ is the arithmetic mean of the same metric (\textit{model performance}, \textit{processing time}, or \textit{expense}) obtained from $50$ runs of the project under one of the other configurations from Table~\ref{env_conf_tbl}.
    \item $P$ is the percentage change between $\overline{m_b}$ and $\overline{m_o}$. Any non-zero value of $P$ indicates the existence of instability in the considering metric.
\end{itemize}

The purpose of this step is to determine, on average across $50$ runs, how much instability can be observed in each of the considering metrics.

\emph{Step 2:} While \emph{Step 1} of our analysis gives us an overall view of instability for each project, in this step, we aim to determine whether or not any observed instability is statistically significant. To determine the statistical significance of any observed instability, we first perform the Mann-Whitney \textit{U} test~\cite{mann1947test} to compare the two distributions. We choose the Mann-Whitney \textit{U} test as the test of statistical significance because this nonparametric test does not assume the data to be normal. We set the level of significance, $\alpha = 0.05$ for this test which represents the traditional $95\%$ confidence level~\cite{chavalarias2016evolution}. Next, we determine the degree of difference, also known as effect size, between the compared distributions using Cliff's delta~\cite{cliff1993dominance}. Cliff's delta, $d$, is bounded between $-1$ and $1$. Based on the value of $d$, the effect size can have one of the following qualitative magnitudes as mentioned in~\cite{khatoonabadi2023understanding,khatoonabadi2023wasted}:

\begin{equation*}
    \text{Effect size}=
    \begin{cases}
        \text{Negligible}, & \text{if         $\lvert d \rvert \leq 0.147$} \\
        \text{Small},      & \text{if $0.147 < \lvert d \rvert \leq 0.33$}  \\
        \text{Medium},     & \text{if $0.33  < \lvert d \rvert \leq 0.474$} \\
        \text{Large},      & \text{if $0.474 < \lvert d \rvert \leq 1$}     \\
    \end{cases}
\end{equation*}

Note that we consider the observed instability between the generated distributions as statistically significant if the Mann-Whitney \textit{U} test returns a \emph{p-value} of less than 0.05 and the effect size obtained from Cliff's delta is not \textit{negligible}.

\medskip

Finally, we categorize the studied projects into three categories based on our two-step analysis previously described: (i) projects that show zero instability, (ii) projects that show non-zero instability which is statistically insignificant, and (iii) projects that show non-zero instability which is statistically significant. While any instability is undesirable, statistically significant instability is even more concerning.

\section{RQ1: (Operating System) How much instability is introduced by changing the operating system in AI-enabled systems?} \label{rq1}
As our first research question, we study instability with respect to operating systems. We perform a comparative analysis among three operating systems: Linux, MacOS, and Windows. Furthermore, we also investigate whether instability can be observed in different distributions of the same operating system. In this case, the comparative analysis is performed among three Linux LTS distributions: Xenial, Bionic, and Focal.

\subsection{Instability with respect to Operating System} \label{os_result}
\subsubsection*{Setup:} To study the effect of operating systems on the instability in AI-enabled systems, we keep the CPU architecture and Python version constant to their baseline values and vary the choice of operating system only.

\subsubsection*{Findings:} 
\begin{figure}
\centering
\includegraphics[width=\columnwidth]{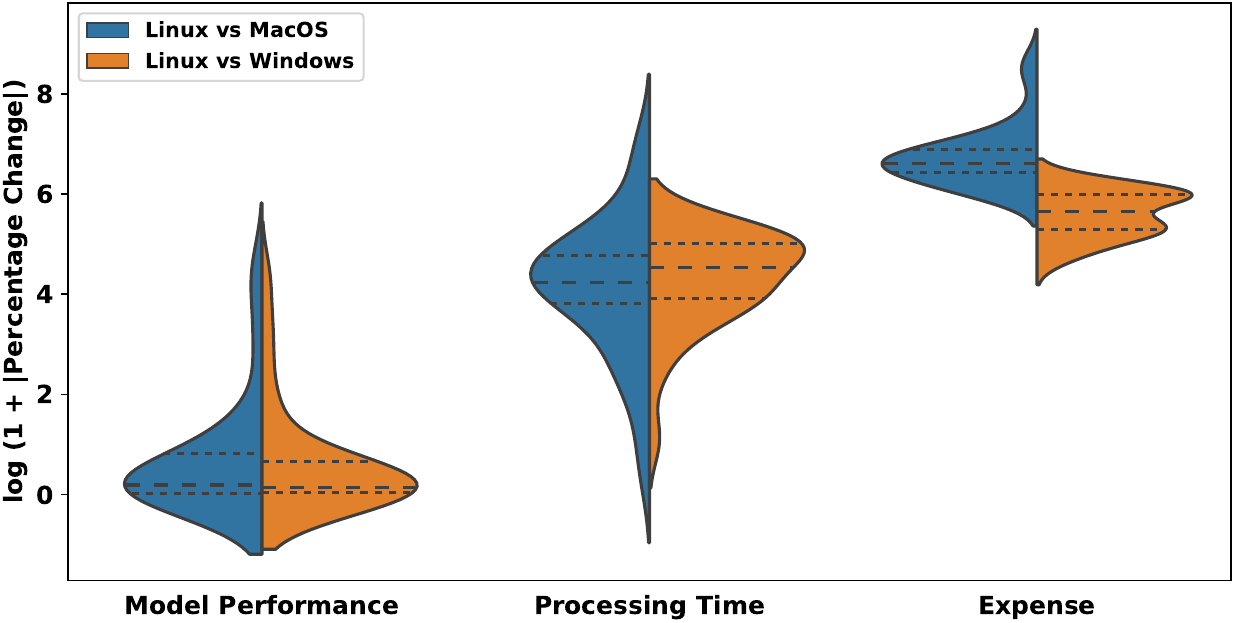}
\caption{Distributions of instability with respect to Operating Systems.}
\label{fig:os_instability}
\end{figure}

We find that the majority of the projects show instability across all three metrics due to operating systems. Figure~\ref{fig:os_instability} shows the distributions of percentage change ($P$) across all the studied projects. Table~\ref{tab:var_cat_os} reports the number of projects falling under different instability categories defined in Section~\ref{var_analysis}. We observe that the majority of the projects show changes in all three metrics due to changes in operating systems. However, only a few projects show instability in model performance with statistical significance. On the other hand, almost all observed instability in processing time and expense is statistically significant. Paying closer attention to the breakdown of effect size for the projects with statistically significant instability, we find that in almost all cases the observed instability is large as shown in Table~\ref{tab:sig_os}. We further find that there is a slight increase in model performance ($1.44\%$) on average when projects are run on MacOS compared to Linux. On the other hand, model performance drops on average by $4.21\%$ when projects are run on Windows. Although a slight increase in model performance may be achieved on MacOS compared to Linux, this will require sacrifice in processing time and expense with MacOS taking $137\%$ longer processing time and costing $1085.47\%$ more money in comparison to Linux.

\textbf{Our findings imply that Linux is a faster and more cost-effective operating system than both MacOS and Windows although MacOS produces slightly better model performance.}

\begin{table*}
\centering
\caption{\label{tab:var_cat_os} Number of projects falling under different instability types due to differences in operating systems.}
\begin{tabular}{|c|c|c|c|}
\hline
\textbf{Metric}                           & \textbf{Instability Type}                      & \textbf{Linux vs MacOS} & \textbf{Linux vs Windows} \\ \hline
\multirow{3}{*}{\textbf{Model Performance}}     & Zero instability                          & 4 (13.33\%)              & 4 (13.33\%)                \\ \cline{2-4} 
                                 & Non-zero but statistically insignificant & 19 (63.33\%)             & 20 (66.67\%)               \\ \cline{2-4} 
                                 & Non-zero and statistically significant    & 7 (23.33\%)              & 6 (20\%)                \\ \hline \hline
\multirow{3}{*}{\textbf{Processing Time}} & Zero instability                          & 0 (0\%)              & 0 (0\%)                \\ \cline{2-4} 
                                 & Non-zero but statistically insignificant & 1 (3.33\%)              & 0 (0\%)                \\ \cline{2-4} 
                                 & Non-zero and statistically significant    & 29 (96.67\%)             & 30 (100\%)               \\ \hline \hline
\multirow{3}{*}{\textbf{Expense}}         & Zero instability                          & 0 (0\%)              & 0 (0\%)                \\ \cline{2-4} 
                                 & Non-zero but statistically insignificant & 0 (0\%)              & 0 (0\%)                \\ \cline{2-4} 
                                 & Non-zero and statistically significant    & 30 (100\%)             & 30 (100\%)               \\ \hline
\end{tabular}
\end{table*}

\begin{table*}
\centering
\caption{\label{tab:sig_os} Breakdown of effect size for projects with statistically significant instability due to different operating systems.}
\begin{tabular}{c|cccc|cccc|}
\cline{2-9}
 &
  \multicolumn{4}{c|}{\textbf{Linux vs MacOS}} &
  \multicolumn{4}{c|}{\textbf{Linux vs Windows}} \\ \cline{2-9} 
 &
  \multicolumn{1}{c|}{Small} &
  \multicolumn{1}{c|}{Medium} &
  \multicolumn{1}{c|}{Large} &
  Total &
  \multicolumn{1}{c|}{Small} &
  \multicolumn{1}{c|}{Medium} &
  \multicolumn{1}{c|}{Large} &
  Total \\ \hline
\multicolumn{1}{|c|}{\textbf{Model Performance}} &
  \multicolumn{1}{c|}{1} &
  \multicolumn{1}{c|}{0} &
  \multicolumn{1}{c|}{6} &
  7 &
  \multicolumn{1}{c|}{0} &
  \multicolumn{1}{c|}{0} &
  \multicolumn{1}{c|}{6} &
  6 \\ \hline
\multicolumn{1}{|c|}{\textbf{Processing Time}} &
  \multicolumn{1}{c|}{0} &
  \multicolumn{1}{c|}{0} &
  \multicolumn{1}{c|}{29} &
  29 &
  \multicolumn{1}{c|}{1} &
  \multicolumn{1}{c|}{0} &
  \multicolumn{1}{c|}{29} &
  30 \\ \hline
\multicolumn{1}{|c|}{\textbf{Expense}} &
  \multicolumn{1}{c|}{0} &
  \multicolumn{1}{c|}{0} &
  \multicolumn{1}{c|}{30} &
  30 &
  \multicolumn{1}{c|}{0} &
  \multicolumn{1}{c|}{0} &
  \multicolumn{1}{c|}{30} &
  30 \\ \hline
\end{tabular}
\end{table*}

\subsection{Instability with respect to Linux Distribution}
\subsubsection*{Setup:} To study whether instability can be observed in different distributions of the same operating system, we vary only the distribution variable in the configuration settings and keep the operating system, Python version, and CPU architecture constant to their baseline values.

\subsubsection*{Findings:} 
\begin{figure}
\centering
\includegraphics[width=\columnwidth]{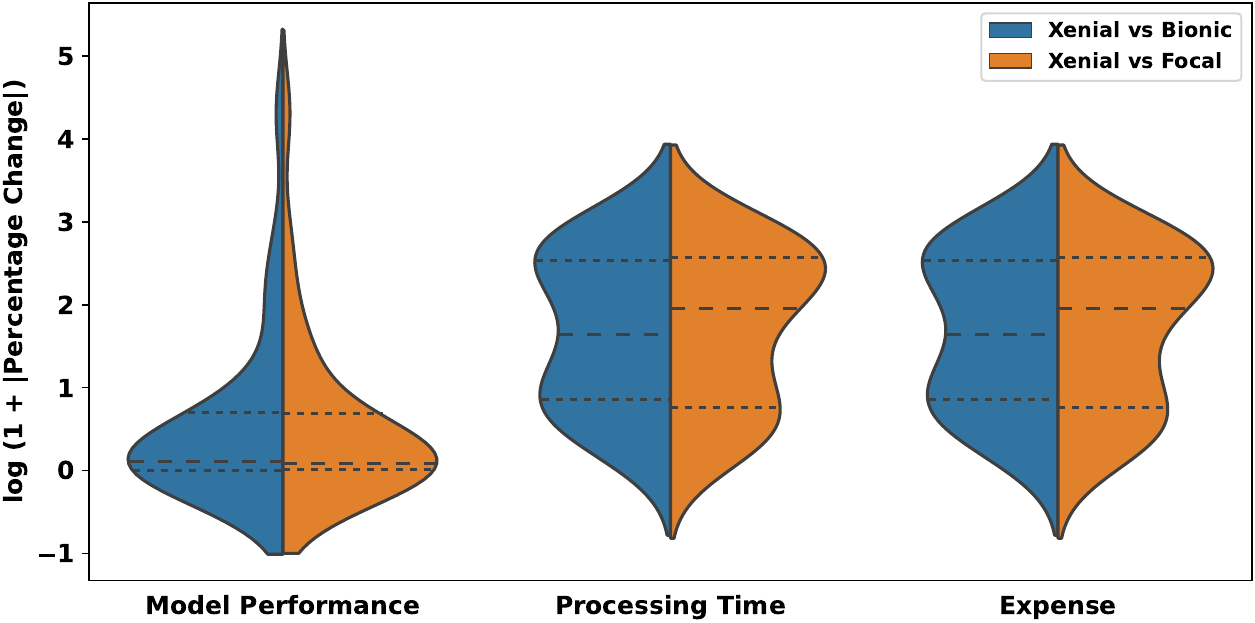}
\caption{Distributions of instability with respect to Linux Distributions.}
\label{fig:dist_instability}
\end{figure}

We find that, similar to operating systems, different Linux distributions also cause varying degrees of instability with respect to all three studied metrics. Figure~\ref{fig:dist_instability} shows the distribution of percentage change ($P$) caused by changes in the Linux distribution across the studied projects. Table~\ref{tab:var_cat_dist} reveals that the majority of the projects show some degree of instability between different distributions of Linux. Although none of the observed instability between Xenial and Bionic is statistically significant in any of the metrics, the observed instability between Xenial and Focal shows a different pattern. Between Xenial and Focal, three projects show a statistically significant instability in model performance whereas $23$ projects show a statistically significant instability in processing time and expense. Most of the observed statistically significant instability is large in terms of effect size as shown in Table~\ref{tab:sig_dist}. On average a slight model performance gain of $2\%$ can be achieved by choosing Focal over Xenial, however, this comes with a $7\%$ increase in processing time and expense. This implies that newer versions of Linux have increased processing time and thus higher expense.

\begin{table*}
\centering
\caption{\label{tab:var_cat_dist} Number of projects falling under different instability types due to differences in Linux distributions.}
\begin{tabular}{|c|c|c|c|}
\hline
\textbf{Metric}                           & \textbf{Instability Type}                      & \textbf{Xenial vs Bionic} & \textbf{Xenial vs Focal} \\ \hline
\multirow{3}{*}{\textbf{Model Performance}}     & Zero instability                          & 7 (23.33\%)                & 6 (20\%)               \\ \cline{2-4} 
                                 & Non-zero but statistically insignificant & 23 (76.67\%)               & 21 (70\%)              \\ \cline{2-4} 
                                 & Non-zero and statistically significant    & 0 (0\%)                & 3 (10\%)               \\ \hline \hline
\multirow{3}{*}{\textbf{Processing Time}} & Zero instability                          & 0 (0\%)                & 0 (0\%)               \\ \cline{2-4} 
                                 & Non-zero but statistically insignificant & 30 (100\%)               & 7 (23.33\%)               \\ \cline{2-4} 
                                 & Non-zero and statistically significant    & 0 (0\%)                & 23 (76.67\%)              \\ \hline \hline
\multirow{3}{*}{\textbf{Expense}}         & Zero instability                          & 0 (0\%)                & 0 (0\%)               \\ \cline{2-4} 
                                 & Non-zero but statistically insignificant & 30 (100\%)               & 7 (23.33\%)               \\ \cline{2-4} 
                                 & Non-zero and statistically significant    & 0 (0\%)                & 23 (76.67\%)              \\ \hline
\end{tabular}
\end{table*}

\begin{table*}
\centering
\caption{\label{tab:sig_dist} Breakdown of effect size for projects with statistically significant instability between different Linux distributions.}
\begin{tabular}{c|cccc|cccc|}
\cline{2-9}
 &
  \multicolumn{4}{c|}{\textbf{Xenial vs Bionic}} &
  \multicolumn{4}{c|}{\textbf{Xenial vs Focal}} \\ \cline{2-9} 
 &
  \multicolumn{1}{c|}{Small} &
  \multicolumn{1}{c|}{Medium} &
  \multicolumn{1}{c|}{Large} &
  Total &
  \multicolumn{1}{c|}{Small} &
  \multicolumn{1}{c|}{Medium} &
  \multicolumn{1}{c|}{Large} &
  Total \\ \hline
\multicolumn{1}{|c|}{\textbf{Model Performance}} &
  \multicolumn{1}{c|}{0} &
  \multicolumn{1}{c|}{0} &
  \multicolumn{1}{c|}{0} &
  0 &
  \multicolumn{1}{c|}{1} &
  \multicolumn{1}{c|}{0} &
  \multicolumn{1}{c|}{2} &
  3 \\ \hline
\multicolumn{1}{|c|}{\textbf{Processing Time}} &
  \multicolumn{1}{c|}{0} &
  \multicolumn{1}{c|}{0} &
  \multicolumn{1}{c|}{0} &
  0 &
  \multicolumn{1}{c|}{4} &
  \multicolumn{1}{c|}{1} &
  \multicolumn{1}{c|}{18} &
  23 \\ \hline
\multicolumn{1}{|c|}{\textbf{Expense}} &
  \multicolumn{1}{c|}{0} &
  \multicolumn{1}{c|}{0} &
  \multicolumn{1}{c|}{0} &
  0 &
  \multicolumn{1}{c|}{4} &
  \multicolumn{1}{c|}{1} &
  \multicolumn{1}{c|}{18} &
  23 \\ \hline
\end{tabular}
\end{table*}

\textbf{Our findings indicate that even though the choice of Linux distribution is unlikely to affect the model performance of AI components significantly, it is very likely to affect the processing time and associated cost of building and running a system.}

\begin{tcolorbox}
    \textbf{RQ1 Findings:} Most projects show significant instability in processing time and cost across operating systems, while only a few exhibit notable differences in model performance.
\end{tcolorbox}

\section{RQ2: (Python Version) How much does changing the Python version introduce instability in AI-enabled systems?} \label{rq2}
\subsubsection*{Setup:} In this research question, we investigate if instability can be observed when different versions of Python are used to run the same system. We study the effect of Python version by keeping the operating system, distribution, and CPU architecture constant to their baseline values and only varying the Python version.

\subsubsection*{Findings:} 
\begin{figure}[t]
\centering
\includegraphics[width=\columnwidth]{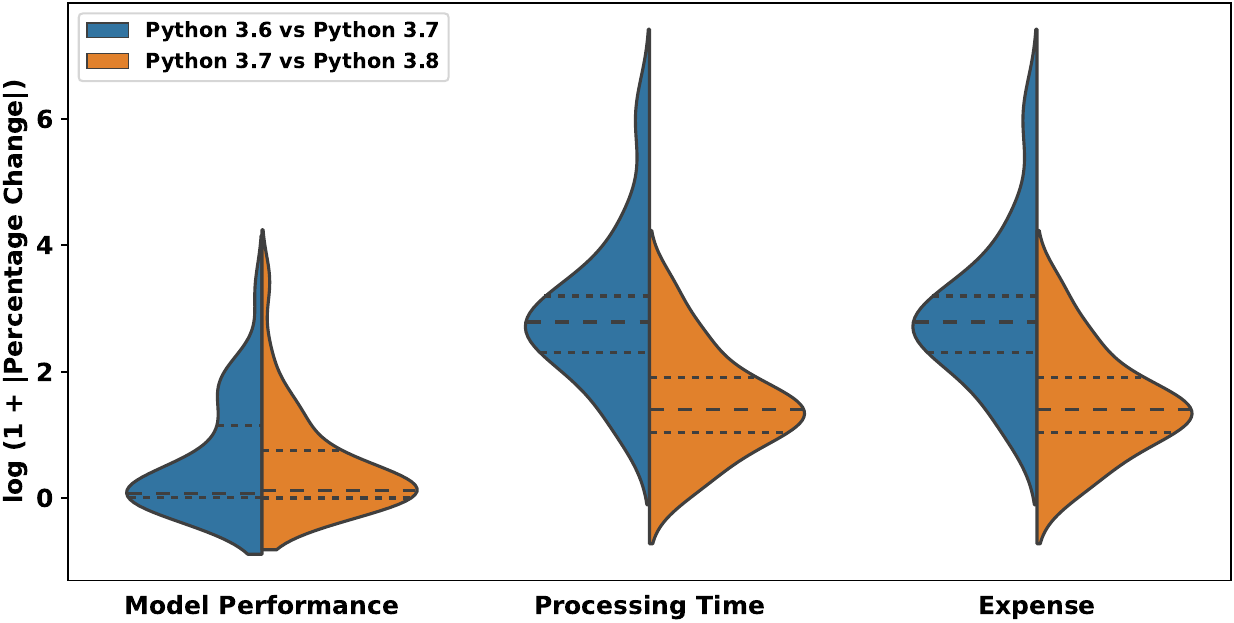}
\caption{Distributions of instability with respect to Python versions.}
\label{fig:pct_diff_py}
\end{figure}

We find that most projects show some degree of instability between Python versions. Figure~\ref{fig:pct_diff_py} shows a similar pattern to the observed instability in RQ1. That said, not all observed instability has statistical significance as shown in Table~\ref{tab:var_cat_py}. Furthermore, Table~\ref{tab:sig_py} reveals that any instability observed between Python 3.6 and Python 3.7 is insignificant in all metrics. On the other hand, five projects with four large effect sizes and one small effect size show significant instability between Python 3.7 and Python 3.8. An even higher degree of instability can be observed in processing time and expense with a total of $24$ projects showing significant instability with $19$ large, two medium, and three small effect sizes. Moreover, choosing Python 3.6 over Python 3.7 causes a $0.52\%$ drop in model performance, and choosing Python 3.8 over Python 3.7 causes a $0.73\%$ drop in model performance on average. To build and run a project it takes $25\%$ longer using Python 3.6 and $5.3\%$ longer using Python 3.8. Expense follows the same pattern as processing time.

\begin{table*}
\centering
\caption{\label{tab:var_cat_py} Number of projects falling under different instability types due to differences in Python versions.}
\begin{tabular}{|c|c|c|c|}
\hline
\textbf{Metric}                           & \textbf{Instability Type}                          & \textbf{Python 3.6 vs Python 3.7} & \textbf{Python 3.7 vs Python 3.8} \\ \hline
\multirow{3}{*}{\textbf{Model Performance}}     & Zero instability                          & 6 (20\%)                        & 6 (20\%)                        \\ \cline{2-4} 
                                 & Non-zero but statistically insignificant & 24 (80\%)                       & 19 (63.33\%)                       \\ \cline{2-4} 
                                 & Non-zero and statistically significant    & 0 (0\%)                        & 5 (16.67\%)                        \\ \hline \hline
\multirow{3}{*}{\textbf{Processing Time}} & Zero instability                          & 0 (0\%)                        & 0 (0\%)                        \\ \cline{2-4} 
                                 & Non-zero but statistically insignificant & 30 (100\%)                       & 6 (20\%)                        \\ \cline{2-4} 
                                 & Non-zero and statistically significant    & 0 (0\%)                        & 24 (80\%)                       \\ \hline \hline
\multirow{3}{*}{\textbf{Expense}}         & Zero instability                          & 0 (0\%)                        & 0 (0\%)                        \\ \cline{2-4} 
                                 & Non-zero but statistically insignificant & 30 (100\%)                       & 6 (20\%)                        \\ \cline{2-4} 
                                 & Non-zero and statistically significant    & 0 (0\%)                        & 24 (80\%)                       \\ \hline
\end{tabular}
\end{table*}

\begin{table*}
\centering
\caption{\label{tab:sig_py} Breakdown of effect size for projects with statistically significant instability between different Python versions.}
\begin{tabular}{c|cccc|cccc|}
\cline{2-9}
 &
  \multicolumn{4}{c|}{\textbf{Python 3.6 vs Python 3.7}} &
  \multicolumn{4}{c|}{\textbf{Python 3.7 vs Python 3.8}} \\ \cline{2-9} 
 &
  \multicolumn{1}{c|}{Small} &
  \multicolumn{1}{c|}{Medium} &
  \multicolumn{1}{c|}{Large} &
  Total &
  \multicolumn{1}{c|}{Small} &
  \multicolumn{1}{c|}{Meidum} &
  \multicolumn{1}{c|}{Large} &
  Total \\ \hline
\multicolumn{1}{|c|}{\textbf{Model Performance}} &
  \multicolumn{1}{c|}{0} &
  \multicolumn{1}{c|}{0} &
  \multicolumn{1}{c|}{0} &
  0 &
  \multicolumn{1}{c|}{1} &
  \multicolumn{1}{c|}{0} &
  \multicolumn{1}{c|}{4} &
  5 \\ \hline
\multicolumn{1}{|c|}{\textbf{Processing Time}} &
  \multicolumn{1}{c|}{0} &
  \multicolumn{1}{c|}{0} &
  \multicolumn{1}{c|}{0} &
  0 &
  \multicolumn{1}{c|}{3} &
  \multicolumn{1}{c|}{2} &
  \multicolumn{1}{c|}{19} &
  24 \\ \hline
\multicolumn{1}{|c|}{\textbf{Expense}} &
  \multicolumn{1}{c|}{0} &
  \multicolumn{1}{c|}{0} &
  \multicolumn{1}{c|}{0} &
  0 &
  \multicolumn{1}{c|}{3} &
  \multicolumn{1}{c|}{2} &
  \multicolumn{1}{c|}{19} &
  24 \\ \hline
\end{tabular}
\end{table*}

\textbf{Our findings show that the choice of Python version can cause instability, likely due to differences in library versions during installation. Newer Python versions may install updated libraries with added features, increasing installation time, while older versions may lead to performance drops and longer processing due to outdated dependencies.}

\begin{tcolorbox}
    \textbf{RQ2 Findings:} Python 3.6 and 3.7 produce identical results across all metrics in the studied projects. However, between Python 3.7 and 3.8, most projects experience significant instability in processing time and costs, with only a few showing instability in model performance.
\end{tcolorbox}

\section{RQ3: (CPU Architecture) How much does changing the CPU architecture introduce instability in AI-enabled systems?} \label{rq3}

\subsubsection*{Setup:} We study the effect of CPU architecture on instability by keeping the operating system, distribution, and Python version constant to their baseline values and only varying CPU architecture configuration.

\subsubsection*{Findings:} 
\begin{figure}
\centering
\includegraphics[width=\columnwidth]{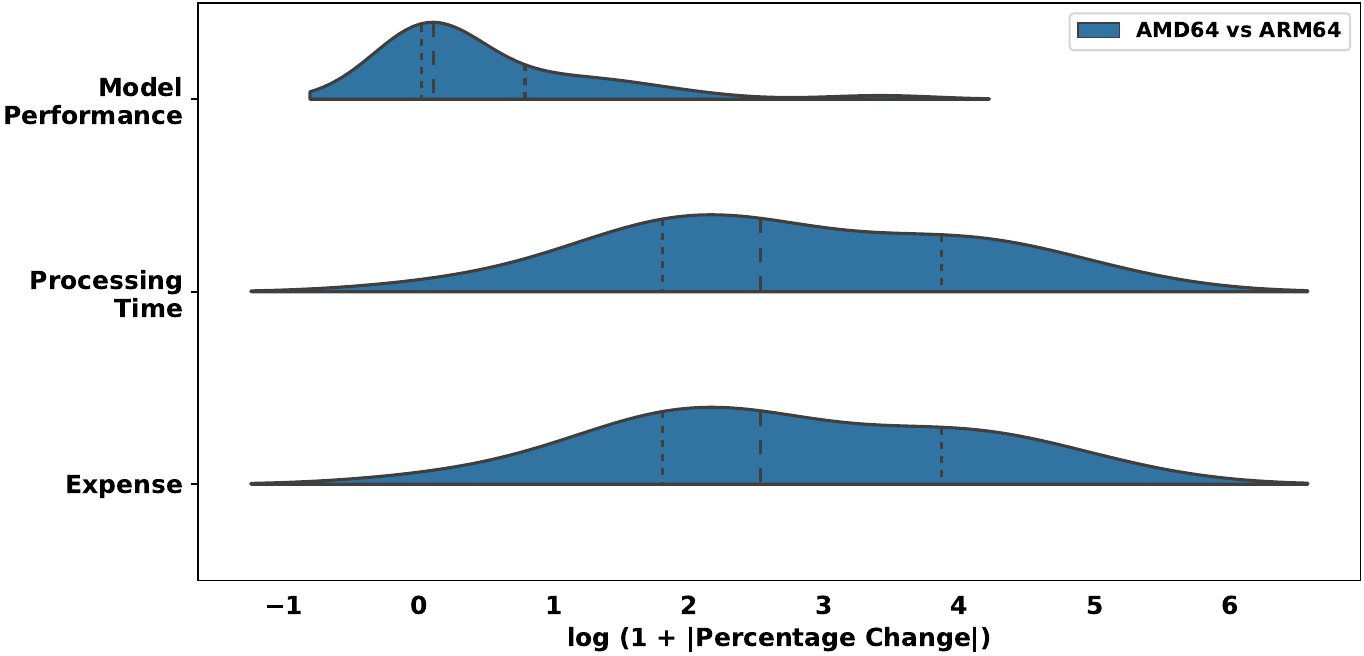}
\caption{Distributions of instability with respect to CPU architectures.}
\label{fig:pct_diff_hw}
\end{figure}

We find that CPU architectures also cause varying degrees of instability in all three metrics. Figure~\ref{fig:pct_diff_hw} summarizes the instability pattern in terms of percentage changes between AMD64 and ARM64 CPU architectures. Similar to the findings from RQ1 and RQ2, most of the observed instability in model performance is insignificant, whereas, in processing time and expense, the observed instability is significant in the majority of the studied projects. Table~\ref{tab:sig_hw} shows that out of six projects with significant instability in model performance, five show a large effect size and one shows a small effect size. In processing time and expense two, one and $25$ projects show small, medium, and large effect sizes respectively among the $28$ projects that differ significantly between AMD64 and ARM64 CPU architectures. In all three metrics, the ARM64 CPU performs poorly compared to the AMD64 CPU with a slight drop of $0.62\%$ in model performance costing $25\%$ more time and money, on average. We conjecture that the observed instability between AMD64 and ARM64 CPU architectures may be due to design differences. ARM64 has a much smaller instruction set compared to AMD64 which might require ARM64 to take longer to perform more complex operations~\cite{bhandarkar1991performance}. AMD64 being the most common CPU architecture has more software support compared to ARM64 CPUs. All these can negatively affect the model performance, processing time, and expense of building and running AI-enabled systems on ARM64 CPUs.

\begin{table*}
\centering
\caption{\label{tab:var_cat_hw} Number of projects falling under different instability types due to differences in CPU architectures.}
\begin{tabular}{|c|c|c|}
\hline
\textbf{Metric}                           & \textbf{Instability Type}                          & \textbf{AMD64 vs ARM64} \\ \hline
\multirow{3}{*}{\textbf{Model Performance}}     & Zero instability                          & 4 (13.33\%)              \\ \cline{2-3} 
                                 & Non-zero but statistically insignificant & 20 (66.67\%)             \\ \cline{2-3} 
                                 & Non-zero and statistically significant    & 6 (20\%)              \\ \hline \hline
\multirow{3}{*}{\textbf{Processing Time}} & Zero instability                          & 0 (0\%)              \\ \cline{2-3} 
                                 & Non-zero but statistically insignificant & 2 (6.67\%)              \\ \cline{2-3} 
                                 & Non-zero and statistically significant    & 28 (93.33\%)             \\ \hline \hline
\multirow{3}{*}{\textbf{Expense}}         & Zero instability                          & 0 (0\%)              \\ \cline{2-3} 
                                 & Non-zero but statistically insignificant & 2 (6.67\%)              \\ \cline{2-3} 
                                 & Non-zero and statistically significant    & 28 (93.33\%)             \\ \hline
\end{tabular}
\end{table*}

\begin{table}
\centering
\caption{\label{tab:sig_hw} Breakdown of effect size for projects with statistically significant instability between different CPU architectures.}
\begin{tabular}{c|cccc|}
\cline{2-5}
                                  & \multicolumn{4}{c|}{\textbf{AMD64 vs ARM64}}                                                           \\ \cline{2-5} 
                                  & \multicolumn{1}{c|}{Small} & \multicolumn{1}{c|}{Medium} & \multicolumn{1}{c|}{Large} & Total \\ \hline
\multicolumn{1}{|c|}{\textbf{Model Performance}} & \multicolumn{1}{c|}{1}     & \multicolumn{1}{c|}{0}      & \multicolumn{1}{c|}{5}     & 6     \\ \hline
\multicolumn{1}{|c|}{\textbf{Processing Time}} & \multicolumn{1}{c|}{2} & \multicolumn{1}{c|}{1} & \multicolumn{1}{c|}{25} & 28 \\ \hline
\multicolumn{1}{|c|}{\textbf{Expense}}     & \multicolumn{1}{c|}{2}     & \multicolumn{1}{c|}{1}      & \multicolumn{1}{c|}{25}    & 28    \\ \hline
\end{tabular}
\end{table}

\textbf{We can draw a similar conclusion for RQ3 to what we observed and concluded in RQ1 and RQ2. Instability in model performance is less common than instability in processing time and associated costs. Therefore, the most optimized hardware configuration can significantly reduce processing time and costs because in the majority of the cases, the observed statistically significant instability is large between AMD64 and ARM64 CPU architectures.}

\begin{tcolorbox}
    \textbf{RQ3 Findings:} CPU architecture has a major impact on processing time and costs in most projects, with a large effect size in these metrtics. While less common, the choice of CPU architecture can occasionally cause instability in model performance.
\end{tcolorbox}

\section{Discussion} \label{discussion}
\subsection{Interpretation}
AI components are becoming a core part of almost all software systems nowadays. Our analysis shows that these systems suffer from instability in all three metrics we studied (\emph{model performance}, \emph{processing time}, and \emph{expense}) although this finding is not consistent across all projects under investigation. Although existing work reported the existence of inconsistent and non-deterministic behavior of AI-enabled systems due to different factors such as choice of frameworks~\cite{zhuang2022randomness,guo2019empirical}, underspecification~\cite{d2022underspecification}, and CPU multithreading~\cite{xiao2021nondeterministic}, ours is the first work to investigate the effect of environment configurations on the systems' stability to the best of our knowledge. We find that the choice of operating system including the distribution of an operating system, version of Python, and CPU architecture indeed introduce instability. The degree of instability differs from project to project. Four projects, namely Seglearn~\cite{seglearn2018}, SMOTE-variants~\cite{smote2018}, pyGLMnet~\cite{pyglmnet2016}, and pymfe~\cite{pymfe2019}, consistently show stable behavior across all three studied metrics with respect to all the configuration settings under investigation. However, our qualitative analysis did not reveal any probable reason behind why this might be happening. Comparing these four projects against other $26$ projects that show some degree of instability in at least one of the studied metrics did not show any distinct feature about them. While performing our qualitative analysis of the codebase of all projects, we did not notice anything different about the implementations of these four projects either, such as setting a value of the {\tt random\_state} setting internally within the implementations forcing an operation to repeatedly produce the same result. Therefore, we couldn't determine why the observed instability varies between projects without further investigation which is out of the scope of the paper and requires its dedicated study.

\subsection{Implications}
Firstly, our results imply that to be able to determine the existence and degree of instability in a project, developers must build and run the project under various configuration settings. Furthermore, our findings indicate that even within a project, not all metrics show an equal degree of instability. Instability is more prominent in processing time and associated costs than the model performance of an AI component. This can hurt a project's development lifecycle given that AI components like ML models need to be retrained frequently because they suffer from model performance decay due to data drift~\cite{nelson2015evaluating} and concept drift~\cite{lu2018learning} over time. If the (re)training of an AI component takes a very long time under a given configuration setting, it can reduce the frequent update of the system and eventually can lead to a model performance drop. Moreover, a longer processing time usually implies higher costs.

Secondly, while instability in processing time and associated expenses impacts only the project internally (such as longer development time and an unnecessary increase in development efforts), the instability in model performance can impact the end-users of the project. This can potentially cause a financial burden for a company. The instability in model performance can be reduced by adhering to the practice of \textit{dev/prod parity} which is an important principle of 12-factor apps~\cite {wiggins2017twelve,hoffman2016beyond}. Therefore, our findings suggest that an AI-enabled system should be built and run on different configuration settings first so that the developers can determine the environment configuration on which the most optimized system can be developed. This step should follow the \textit{dev/prod} parity principle to ensure the stability of the system. Based on existing literature~\cite{john2021architecting}, it is not yet a common practice in the industry to build and run a system on multiple different configuration settings. Our findings imply that determining the best configuration setting with respect to the metric(s) of interest should be an important step in the development workflow.

\section{Threats to Validity} \label{threats}
\subsubsection*{Internal Validity:} There are projects in our dataset that are developed for more than one task. However, we only run one example task as part of our example script for each project to perform an analysis of the outputs. It is entirely possible that the example tasks may not represent the actual degree of instability associated with the project. To mitigate this issue, we only choose an example task that is part of the official documentation of each project with a naive assumption that the developers would choose those examples in such a way that they are a true representation of the overall functionality and model performance of the project.

\subsubsection*{External Validity:} Firstly, we cannot guarantee the generalizability of our findings. We chose a finite set of configuration variables with a finite set of possible values for each of the variables under study. However, we acknowledge that there are other options for each of these variables that we do not investigate. For example, Linux has many distributions other than Xenial, Bionic, and Focal. Python has many other versions besides the ones we studied. Therefore, we do not claim that our findings can be generalized beyond what we investigated. The reason behind limiting our choices of options for the configuration variables is the amount of time and money required to run experiments in Travis CI. Furthermore, for each new configuration setting, we would have had to run $50$ iterations because of our experimental design. Doing so was not practically feasible due to constraints on time and money.

Secondly, we acknowledge that the dataset with $30$ projects falls at the smaller end for quantitative experiments. As mentioned in Section~\ref{data}, we started with a much bigger set of projects, however, the biggest requirement of our experimental design was that a project needed to successfully build and run on all configurations listed in Table~\ref{env_conf_tbl}. Only $30$ projects met that requirement. That said, these $30$ projects represent a diverse set of applications in terms of size, popularity, and activity as evident from Table~\ref{tab:project-overview}. Furthermore, our manual analysis revealed that these projects represent the implementation of diverse ML algorithms and applications including natural language processing (such as Doc2Vec~\cite{doc2vec2018}), time-series analysis (such as Seglearn~\cite{seglearn2018}), and recommender systems (such as Spotlight~\cite{spotlight2017}).

\section{Related Work} \label{related-work}
\subsubsection*{Nondeterminism in AI:} Uncertain nature of AI components has been a topic of research in the domain of AI for quite some time. It has gained more traction with the popularity of deep learning systems. Most of the existing works on the non-deterministic nature of AI components focus on deep learning systems. For example, Zhuang~\etal~\cite{zhuang2022randomness, pham2020problems} studied the uncertain nature of training deep learning models. They reported that the choice of tools can affect the behavior of an AI component which can potentially affect AI safety. Guo~\etal~\cite{guo2019empirical} performed an empirical study on the development and deployment of deep learning solutions. They reported that frameworks and platforms can cause the model performance of a system to decline. Crane~\cite{crane2018questionable} studied the challenges in the reproducibility of published results. This study reported that the consistent use of random seeds can help mitigate the issue of lack of reproducibility. Xiao~\etal~\cite{xiao2021nondeterministic} reported the impact of CPU multithreading and how it impacts the training of deep learning systems.

\subsubsection*{Instability in Software:} Instability in software systems has been explored in various contexts, including cloud infrastructure, system growth, and reproducibility. Some studies focused on detecting instability or proposing design practices to reduce it~\cite{datskova2017detection,davis2017realizing}. Others found that instability can increase with codebase size~\cite{mresa2021assessing}, or that stable domain abstractions help maintain structural consistency~\cite{maisikeli2018measuring}. Additionally, researchers have identified frequently modified system regions as unstable and proposed methods to prioritize them for restructuring~\cite{bevan2003identification,anda2008variability}.

\subsubsection*{AI-components in Software:} Many recent studies have investigated the pros and cons of having AI components embedded in software systems. Masuda~\etal~\cite{masuda2018survey} described practices for the evaluation and improvement of the software quality of ML applications. Washizaki~\etal~\cite{washizaki2019studying} proposed architecture and design patterns for ML systems. An extensive study on testing ML applications was performed in~\cite{zhang2020machine} by Zhang~\etal. Scully~\etal~\cite{sculley2015hidden} studied hidden technical debt in ML systems whereas Obrien~\etal~\cite{obrien202223} studied self-admitted technical debts in ML software.

\smallskip

\textit{Our work is different from the above studies in that ours is the first study to quantify the degree of instability in AI-enabled systems in terms of model performance, processing time, and expense as a result of changes in configuration settings of three key environment variables (operating system, Python version, and CPU architecture).}

\section{Conclusion and Future Work} \label{conclusion}
In this paper, we investigated how AI-enabled software systems show instability in terms of three metrics: model performance, processing time, and expense of building and running a system. We performed our study with respect to three environment variables, namely operating systems including the distributions of an operating system, Python version, and CPU architecture. Our findings indicate that although a majority of the projects show some degree of instability, the degrees vary from project to project. The instability is more statistically significant for processing time and expense than the model performance of an AI component. Because the observed instability patterns vary from project to project, we conclude that to serve the end users the most accurate AI solutions, it is crucial to run and test the AI components in different environment configurations. This practice can facilitate the identification of the environment where the \textbf{most optimized} system can be built which should follow adherence to the \textit{dev/prod parity} principle to obtain the \textbf{most stable} system. We acknowledge that predicting instability without testing different configurations could save time and effort, making it a valuable topic for future research. Another potential study could explore the causes of observed instability, which we leave for future work.

\bibliographystyle{ACM-Reference-Format}
\bibliography{bibliography}

\end{document}